\documentclass[12pt]{article}
\usepackage[dvips]{color}
\usepackage{amsmath}
\usepackage{graphicx}
\usepackage{subfigure}
\usepackage{epsfig}
\usepackage{amsmath}
\usepackage{cite}
\usepackage{color}
\usepackage{subfigure}

\begin{document}
\begin{center}
\Large{\bf Exploring the Parameter Space of Inflation Model on the Brane and its Compatibility with the Swampland Conjectures}\\
\small \vspace{1cm}
 {\bf Saeed Noori Gashti$^{\dagger}$\footnote {Email:~~~saeed.noorigashti@stu.umz.ac.ir}}\quad
 {\bf Mohammad Reza Alipour$^{\star}$\footnote {Email:~~~mr.alipour@stu.umz.ac.ir}},\quad
and\quad{\bf Mohammad Ali S. Afshar $^{\star}$\footnote {Email:~~~m.a.s.afshar@gmail.com}}\\
\vspace{0.2cm}$^{\star}${Department of Physics, Faculty of Basic
Sciences,\\
University of Mazandaran
P. O. Box 47416-95447, Babolsar, Iran}\\
\vspace{0.5cm}$^{\dagger}${School of Physics, Damghan University, Damghan, 3671641167, Iran.}\\
\small \vspace{1cm}
\end{center}
\begin{abstract}
In recent years, inflationary models have been studied from different perspectives using different conditions. Now, in this paper, we want to investigate the inverse monomial Inflation$(IMI)$ on the brane. We limit our results to the potentials of the form $\frac{M^{4+\sigma}}{\phi^{\sigma}}$. We calculate some cosmological parameters and then we investigate the satisfaction of the model with some conjectures of the swampland program. We will check the model's compatibility with the refined dS swampland conjecture(RDSSC), further refining the de Sitter swampland conjecture (FRSDC), scalar weak gravity (SWGC), and strong scalar weak gravity conjecture (SSWGC). Despite the incompatibility with (DSSC), We find a specific region of compatibility with other ones which means it satisfied the (FRDSSC)  for example with (a=0.982873,\hspace{0.1cm} b=0.017127,\hspace{0.1cm} q=2.2). Also, the model is compatible with the (SWGC) with the ($\phi\leq \sigma+2$) condition and consistent with the (SSWGC) with the ($\phi \leq \sqrt{(\sigma+2)(\sigma+1)}$) condition. Since inverse monomial Inflation(IMI) on the brane could satisfy some swampland conjecture simultaneously, it has the potential to be a “real” inflation model of the universe.
\end{abstract}
\section{Introduction}
Recently the weak gravity conjecture\cite{101,102,103,104,105,106,107,108,108'} as an important conjecture in the swampland program have been studied in various concepts of cosmology such as inflation that One of the central idea of modern cosmology\cite{777,7777,77777}, black hole physics, etc. In this scenario gravity is the weakest force. The weak gravity conjecture is formulated in flat space-time. On the other hand, we know that to unite the four fundamental forces in nature, one uses a theory called string theory, which has two open and closed strings. the open and closed strings are representations of the gauge and gravitational theory, respectively. At high energy levels, with respect to the fact that string is a compatible theory, it has an area called landscape (an area that is a set of effective theories compatible with quantum gravity). But one of the main problems with this theory is that it is not visible in the laboratory, so we need a method for testing it. At low energy levels, theories that are compatible with gravity can only be seen in small areas (landscape) which are surrounded by a large area called a swampland (an area where the set of incompatible theories with quantum gravity(QG)). We also know at low energy levels, whenever we talk about quantum gravity, we have to use effective low-energy theories. As a result, researchers investigated some concepts such as inflationary models by using the conjectures of the swampland program that you can see in Ref.s\cite{1,1',2,2',3,4,5,6,7,8,9,10,11,12}. In recent years, it has been suggested that if we have a compatible theory with QG, it must have some criteria in its background or in other words, it must satisfy some conjectures. These criteria represent a series of restrictions that are challenged with cosmological concepts\cite{109,110,111,112,113,114,115,116,117,118,119,120,121,122,123,124,125,126,127,13,128,129,14,15,16,17} They also cross-reference their findings with the most recent observational data. So, we will have,
\begin{equation}\label{1}
S=\int d^{4}x\sqrt{-g}\bigg(-\frac{1}{2}M_{p}^{2}R+\frac{1}{2}g_{\mu\nu}W^{ij}\partial^{\mu}\phi_{i}\partial^{\nu}\phi_{j}-V\bigg),
\end{equation}
Where $\phi_{i}$ represents field and $W^{ij}$ denotes the field space metric. We analyze the model in terms of the action (1). This examination is facilitated by the DSSC and RDSSC, with $M_p=1$\cite{18,19},
\begin{equation}\label{2}
\frac{\Delta\phi}{M_{pl}}<\mathcal{O}(1),\hspace{0.5cm} \frac{\nabla V}{V}\geq c_1,\hspace{0.5cm} \frac{min(\nabla_i\nabla_j V)}{V}\leq -c_2,                                                                                                    \end{equation}
Considering $c_1$ and $c_{2}$ as constant quantities, researchers derived a novel conjecture by combining the de Sitter and refined de Sitter conjectures. This new conjecture, named the 'further refining de Sitter swampland conjecture,
\begin{equation}\label{3}
(\frac{\nabla V}{V})^{q}-a\frac{min(\nabla_i\nabla_j V)}{V}\geq b,
\end{equation}
where $a =1-b$, $q >2$, a, $b > 0$. a, b, and q are free parameters. Additionally, one can obtain,
\begin{equation}\label{4}
\epsilon_{V}=\frac{1}{2}(\frac{\nabla V}{V})^{2}=\frac{1}{2}F_{1}^{2},\hspace{0.5cm}
\eta_{V}=\frac{min(\nabla_i\nabla_j V)}{V}=F_{2}.
\end{equation}
Thus, based on above equations, we can express the RDSC as follows:
\begin{equation}\label{5}
F_{1}^{q}-aF_{2}\geq1-a.
\end{equation}
Also, we will have,
\begin{equation}\label{6}
F_{1}=\sqrt{2\epsilon_{V}}=\sqrt{\frac{r}{8}},\hspace{0.5cm}F_2=\eta_{V}=\frac{n_{s}+\frac{3r}{8}-1}{2}.
\end{equation}
WGC is a conjecture within the swampland program. It posits that gravity is the weakest force. Additionally, Palti extended this conjecture to propose that gravity is even weaker than the force exerted by scalar fields. This extension is known as SWGC. According to the SWGC, when a particle of mass \(m\) is coupled to a scalar field, assuming \(m^2 = \frac{{\partial^2 V}}{{\partial \phi^2}}\), the following condition holds:
\begin{equation}\label{7}
(V^{(3)})^2\geq(V^{(2)})^2,
\end{equation}
where the number in parentheses represents the order of the derivative with respect to $(\phi)$. Additionally, there exists a stronger version of the SWGC:
\begin{equation}\label{8}
2(V^{(3)})^2-V^{(2)}V^{(4)}\geq(V^{(2)})^2.
\end{equation}
Some points should be summarized regarding the concepts raised as well as the study of the inflationary model on brane according to the condition of the swampland. Therefore, the number of e-folds for the slow-roll inflation of a single field is introduced as follows.
\begin{equation}\label{9}
N=\frac{1}{M_{pl}^{2}}\int\frac{V}{V'}d\phi\approx\frac{\frac{\Delta\phi}{M_{pl}}}{M_{pl}(\frac{V'}{V})},
\end{equation}
where V and $V'$ are potential and its derivative with respect to $\phi$. as you see, somehow in fact the first criteria give us a number of e-folds. This states that it may not possible to have large e-folds. Also, the initial density perturbation resulting from inflationary fluctuations seems to be the second limitation incompatibility with the tensor-to-scalar ratio. Recently, various methods and mechanisms have been proposed to circumvent these constraints in single-field inflation, including the curvaton-like mechanism \cite{20,21}. It is generally stated that quintessential brane inflation correspondent with the swampland criterion, but there are some drawbacks to the initial conditions of non-Bunch-Davies components. It can also be stated that different inflationary models such as warm inflation are consistent with this criteria\cite {22}. All of the above concepts were the motivation which this paper to be organized as follows.\\
In Section 2, we introduce inflation on the brane. In Section 3, we investigate the Inverse Monomial Inflation $(IMI)$ according to the mechanism proposed in section 2 and we calculate some cosmological parameters such as the tensor-to-scalar ratio, the scalar spectrum index and etc. In Section 4, we challenge our model with some conjectures of the swampland program such as (RDSSC), (FRDSSC), (SWGC) and (SSWGC). Then we study the satisfaction of our model with these conjectures, and finally, we express the results in Section 5

\section{The Inflationary Models on the Brane}
In studies related to the inflation model on the brane, we must have a series of modified relations. In fact, in a scenario related to braneworld that our 4-D world is actually a 3-brane that is located in the higher dimensions bulk. We present an overview of 3-brane cosmology within a five-dimensional space-time framework. A detailed analysis is conducted. We assume the universe is filled with a perfect fluid characterized by energy density \( \rho(t) \) and pressure \( p(t) \). In the braneworld cosmological model, the metric induced on the brane is a spatially flat Friedmann–Robertson–Walker (FRW) metric. The Friedmann equation on the brane, which describes the time evolution of \( a(t) \), is given by\cite{5000,5001,5000',5000''},
\begin{equation*}\label{10}
H^{2}=\frac{\Lambda_4}{3}+(\frac{8\pi}{3M_4^2})\rho+(\frac{4\pi}{3M_5^2})^2\rho^2+\frac{\epsilon}{a^4},
\end{equation*}
Here, \( H(t) \) is the Hubble parameter, \( a \) is the scale factor, \( \Lambda_4 \) is the four-dimensional cosmological constant, and the final term represents the influence of bulk gravitons on the brane. The term \( \epsilon/a^4 \) is referred to as dark radiation, as it behaves similarly to radiation. However, during inflation, this term is rapidly diluted and can be neglected. Therefore, according to equations, we are faced with a series of modifications, for example, with respect to \cite{23,24,25,26,27,28,29,30,300}, Friedmann modified equations are expressed as follows.
\begin{equation}\label{10}
H^{2}=\frac{1}{3M_{pl}}\rho(1+\frac{\rho}{2\lambda}),
\end{equation}
Assuming that inflation quickly renders any dark radiation term negligible, this simplifies to the standard Friedmann equation for $( \rho \ll \lambda )$. If the universe is dominated by a scalar field $( \phi )$ with potential $( V(\phi) )$, we can apply the slow-roll approximation to express it as follows\cite{1000,1001,1002}:
\begin{equation*}\label{10}
H^{2}=\frac{1}{3M_{pl}}V(\phi)(1+\frac{V(\phi)}{2\lambda}).
\end{equation*}
The scalar field follows the standard slow-roll equation,
\begin{equation*}\label{10}
3H\dot{\phi}\simeq-V'(\phi).
\end{equation*}
Given the above relation $\lambda$, provides a relations between $M_{4}$ and $M_{5}$. The relations are as follows.
\begin{equation}\label{11}
M_{4}=\sqrt{\frac{3}{4\pi}}(\frac{M_{5}^{2}}{\sqrt{\lambda}})M_{5}
\end{equation}
where $M_{4}$ and $M_{5}$ are the 4-dimensional Planck scale and 5-dimensional Planck scale, respectively. $M_{pl}=\frac{M_{4}}{\sqrt{8\pi}}$ is the reduce Planck and the constraint $M_{5}\geq10^{5}TeV$. If the universe is primarily governed by a scalar field $\phi$ with a potential V($\phi$), we can employ the slow-roll approximation to $H^{2}=8\pi V(\phi)\big(1+V(\phi)/2\lambda\big)/3M_{4}^{2} $. So the scalar field obeys the usual slow-roll equation $3H\dot{\phi}\simeq-V'(\phi)$. It should be noted that in the concepts related to inflation, there exist a series of parameters for calculations, including slow-roll parameters. the modified slow-roll parameters\cite{27} are as follows,
\begin{equation}\label{12}
\epsilon\equiv\frac{1}{2\kappa^2}(\frac{V'}{V})^{2}\frac{1}{(1+\frac{V}{2\lambda})^{2}}(1+\frac{V}{\lambda})
\end{equation}
\begin{equation}\label{13}
\eta\equiv(\frac{V''}{V})(\frac{1}{1+\frac{V}{2\lambda}})
\end{equation}
where $\kappa^2= 8\pi G=8\pi/M_{4}^{2}$ and $ M_{4}=1.22\times10^{28}$ eV. Inflation concludes when either of the two slow parameters reaches one. In this scenario, calculating the number of e-foldings from the beginning to the end of inflation becomes straightforward. The expansion amount, expressed in e-foldings, is given by,
\begin{equation}\label{14}
N=\int_{\phi_{e}}^{\phi_{i}}(\frac{V}{V'})(1+\frac{V}{2\lambda})d\phi
\end{equation}
Also, the spectrum (PR) are given by,
\begin{equation}\label{15}
P_{R}=\frac{1}{12\pi^{2}}\frac{V^{3}}{V'^{2}}(1+\frac{V}{2\lambda})^{3}
\end{equation}
The spectrum of scalar perturbations is written as,
\begin{equation}\label{16}
A^{2}_{S}=\frac{4}{25}\frac{H^{2}}{\dot{\phi}^{2}}\big(\frac{H}{2\pi}\big)^{2}\simeq\frac{512\pi}{75M_{4}^{6}}\frac{(V(\phi))^{3}}{(V(\phi))^{2}}\big(1+\frac{V(\phi)}{2\lambda})^{3}
\end{equation}
Within the framework of the slow-roll approximation, the scale dependence of the scalar power spectrum is governed by the scalar spectral index, which follows the standard relation,
\begin{equation}\label{17}
n_{s}-1=\frac{d\ln A_{S}^{2}}{d\ln k}\simeq+2\eta-6\epsilon
\end{equation}
In the high-energy approximation, both $\epsilon$ and $\eta$ are suppressed. Within the braneworld context, we find that \( n_s \) is very close to 1, resulting in the Harrison–Zel’dovich spectrum (\( n_s = 1 \)). The running of the spectral index (\( \alpha_s \)) is also presented, and it is possible to define the variation of the spectral index,
\begin{equation}\label{18}
\alpha_s=\frac{dn_{s}}{d\ln k}\bigg|_{k=aH}\simeq-24\epsilon^{2}+16\epsilon\eta-2\zeta^{2}
\end{equation}
Also, the amplitude of tensor perturbations is as follows,
\begin{equation}\label{19}
A^{2}_{T}=\frac{32}{75M_{4}^{4}}\bigg(V(\phi)\bigg[1+\frac{V(\phi)}{2\lambda}\bigg]\bigg)\bigg|_{k=aH}
\end{equation}
So the tensor spectral index is given by,
\begin{equation}\label{20}
n_{T}=\frac{d\ln A_{T}^{2}}{d\ln k}\bigg|_{k=aH}\simeq-2\epsilon
\end{equation}
Consequently, the tensor-to-scalar ratio is as follows
\begin{equation}\label{21}
r=\frac{A_{T}^{2}}{A_{S}^{2}}=\bigg(\epsilon\bigg[\frac{1}{1+\frac{V(\phi)}{\lambda}}\bigg]\bigg)\bigg|_{k=aH}
\end{equation}
According to all the contents as well as the parameters introduced above in the next section, we introduce our inflationary model. Then we will calculate the cosmological parameters mentioned above and we will challenge the model with swampland conjectures.

\section{IMI on the Brane}
In this section, we want to investigate the $(IMI)$ inflation model\cite{28,a,b}, according to the concepts and parameters mentioned in the previous section. This inflation model is usually denoted by an inverse monomial potential. We limit our results to the potentials of the following form. This potential has been extensively examined about quintessence in brane inflation and tachyonic inflation. A notable finding from these studies is that inflation occurs when \(\sigma > 2\)\cite{b',b''}.
\begin{equation}\label{22}
V(\phi)=\frac{M^{4+\sigma}}{\phi^{\sigma}},
\end{equation}
where $M$ and $\sigma$ are constants. The corresponding model with an exponential potential was investigated in\cite{a,b} and in the context of tachyonic inflation\cite{c}, an intriguing finding by Huey and Lidsey\cite{a,b} is that inflation occurs for values of $\sigma$ greater than 2. At high energies, the standard slow-roll parameters undergo modification due to the quadratic correction in the Friedmann equation. In the following path, we use the condition $\frac{V}{\lambda}\gg1$. Therefore, the effects related to brane will be significant, this condition can show its upper bound of $(\lambda)$. Now according to equations (12), and (13) one can be obtained.
\begin{equation}\label{23}
\epsilon=\frac{1}{2\kappa^{2}}\sigma^2\phi^{\sigma+2}\bigg[\frac{4\lambda(\phi^{\sigma}+M^{4+\sigma})}{(2\lambda\phi^{\sigma}+M^{4+\sigma})^{2}}\bigg]
\end{equation}
\begin{equation}\label{24}
\eta=\frac{1}{2\kappa^{2}}\sigma(\sigma+1)\phi^{\sigma-2}\bigg[\frac{4\lambda}{2\lambda\phi^{\sigma}+M^{4+\sigma}}\bigg]
\end{equation}
With respect to the approximation $\lambda\ll V(\phi)$, and  slow-roll brane inflation, we will have,
\begin{equation}\label{25}
\epsilon\simeq\frac{4\lambda\sigma^{2}}{2\kappa^{2}}\frac{\phi^{\sigma-2}}{M^{4-\sigma}}
\end{equation}
\begin{equation}\label{26}
\eta=\frac{2\lambda\sigma(\sigma+1)\phi^{\sigma-2}}{\kappa^{2}M^{4+\sigma}}
\end{equation}
and
\begin{equation}\label{27}
\zeta^2\simeq \frac{\lambda^{2}}{2\kappa^{2}}\frac{V'V''}{V^4}\simeq\frac{\lambda^{2}}{2\kappa^{2}}\frac{\sigma^{2}(\sigma+1)(\sigma+2)}{M^{8+2\sigma}}\phi^{2\sigma-4}
\end{equation}
As you know, inflation ends at $\phi=\phi_{e}$ if one of the slow-roll parameters is the order of one, that is, in other words ($\epsilon=1$) or ($\eta=1$). according to $\epsilon=1$, one can calculate,
\begin{equation}\label{28}
\phi_{e}=\bigg(\frac{2\kappa^{2}}{4\lambda}\frac{M^{4+\sigma}}{\sigma^{2}}\bigg)^{\frac{1}{\sigma-2}}
\end{equation}
and
\begin{equation}\label{29}
V_{e}=\frac{1}{\big(\frac{\kappa^2}{2\lambda\sigma^2}\big)^{\frac{\sigma}{\sigma-1}}\big(M^{\frac{4+\sigma}{\sigma-1}}\big)}
\end{equation}
so, with respect to equation (14) and $\lambda\ll V$, we calculated the number of e-folds which is given by,
\begin{equation}\label{30}
N=\frac{\kappa^{2}M^{\frac{2(4+\sigma)}{\sigma}}}{2\lambda\sigma(2-\sigma)}\bigg(V_{e}^{\frac{\sigma-2}{\sigma}}-V_{i}^{\frac{\sigma-2}{\sigma}}\bigg)
\end{equation}
\begin{equation}\label{31}
N=\frac{\kappa^{2}M^{4+\sigma}}{2\lambda\sigma(2-\sigma)}\bigg(\phi_{e}^{2-\sigma}-\phi_{i}^{2-\sigma}\bigg)
\end{equation}
The amplitude of the scalar and tensor perturbations made during inflation are as follows,
\begin{equation}\label{32}
A_{S}^{2}=\frac{\kappa^{6}}{75\pi^{2}}\frac{V^{3}}{V'^{2}}\bigg(\frac{2\lambda+V}{2\lambda}\bigg)^3
\end{equation}
\begin{equation}\label{33}
A_{T}^{2}=\frac{\kappa^{4}}{150\pi^{2}}V\bigg(\frac{2\lambda+V}{2\lambda}\bigg)
\end{equation}
The amplitude of the scalar perturbation is approximated at the very high energy as follows,
\begin{equation}\label{34}
A_{S}^{2}=\frac{\kappa^{6}}{600\pi^{2}\lambda^3}\frac{V^{6}}{V'^{2}}=\frac{\kappa^{6}}{600\pi^{2}\lambda^3}\bigg(\frac{M^{4+\sigma}}{\phi^{2\sigma-1}}\bigg)^2
\end{equation}
So the above equation in terms of N and $\sigma$ is as follows,
\begin{equation}\label{35}
A_{S}^{2}=\frac{\bigg(\sigma-(\sigma-2)N\bigg)^2}{600\pi^2 \lambda\kappa^2}
\end{equation}
The scale-dependence of the perturbations is given by the mean of the spectral indices,
\begin{equation}\label{36}
n_{S}-1=\frac{d\ln A_{S}^{2}}{d\ln k}\simeq-6\epsilon+2\eta=\frac{4\lambda\sigma(1-2\sigma)}{\kappa^{2}}\frac{\phi^{\sigma-2}}{M^{4+\sigma}}
\end{equation}
\begin{equation}\label{37}
n_{T}=\frac{d\ln A_{T}^{2}}{d\ln k}=\frac{4\lambda\sigma(\sigma+1)}{\kappa^{2}}\frac{\phi^{\sigma-2}}{M^{4+\sigma}}
\end{equation}
At The horizon crossing, the slow parameter are calculated as,
\begin{equation}\label{38}
\epsilon\simeq\frac{\sigma}{\sigma-(\sigma+2)N}
\end{equation}
\begin{equation}\label{39}
\eta\simeq\frac{\sigma+1}{\sigma-(\sigma+2)N}
\end{equation}
\begin{equation}\label{40}
\zeta\simeq\frac{(\sigma+1)(\sigma+2)}{(\sigma-(\sigma+2)N)^2}
\end{equation}
So from the above equations, we can calculate,
\begin{equation}\label{41}
n_{s}-1\simeq\frac{2-4\sigma}{\sigma-(\sigma+2)N}
\end{equation}
\begin{equation}\label{42}
n_{T}\simeq-\frac{2\sigma}{\sigma-(\sigma+2)N}
\end{equation}
Also the running spectral index which is given by,
\begin{equation}\label{43}
\alpha_{S}\simeq\frac{-24\sigma^2+16\sigma(\sigma+1)-2(\sigma+1)(\sigma+2)}{(\sigma-(\sigma+2)N)^2}
\end{equation}
Also the tensor-to-scalar ratio is obtained with respect to N and $\sigma$,
\begin{equation}\label{44}
r=\frac{\big(2\sigma(\sigma-(2-\sigma)N)\big)^{\frac{2}{2+\sigma}}}{-2\sigma\big(\sigma^2 -(\sigma+2)^2 N^2\big)}
\end{equation}
After calculating the modified parameters for the $(IMI)$ inflation model according to the above points mentioned, we will now determine the range of each of the parameters. To check our inflationary model, according to the different conjectures of the swampland program, we must obtain different values of cosmological parameters, including the scalar spectrum index and the tensor-to-scalar ratio.
Therefore, according to $(N=55, \hspace{0.1cm}\sigma=3.3)$, the $(n_{s}, r)$ will be $(0.961138,\hspace{0.1cm} 0.0000185195)$, also, for $(N=60, \hspace{0.1cm}\sigma=3.3)$, we have $(0.964411, \hspace{0.1cm}0.0000160583)$, and for the $(N=70, \hspace{0.1cm}\sigma=3.3)$, we can calculate $(0.96954,\hspace{0.1cm} 0.0000124768)$, respectively. As it is clear, we can calculate different values for two of the most important cosmological parameters by setting different values of $(N, \sigma)$. in the following, we consider the $(N=60, \hspace{0.1cm}\sigma=3.3)$ for the check of swampland conjectures. The parameter \( r \) represents the tensor-to-scalar ratio, which measures the relative contribution of gravitational waves (tensor perturbations) to density fluctuations (scalar perturbations) in the early universe. A very small value of \( r \) is considered advantageous. Current observations, particularly from the Planck satellite, suggest that \( r \) is very small. Models of inflation that predict a small \( r \) are therefore more consistent with these observations.
A small \( r \) implies that the inflationary period was very efficient at smoothing out any initial irregularities and flattening the universe. This helps explain why the universe appears so homogeneous and isotropic on large scales. Inflationary models that predict a small \( r \) often have other testable predictions that can be compared with observations. This makes these models more robust and scientifically valuable.
\section{Swampland Conjecture $\&$ IMI on Brane}
In this section, we investigate the implications of the IMI  on the brane in light of swampland conjectures, specifically the FRDSSC , SWGC, and SSWGC. Additionally, we analyze how well the model satisfies each of these conjectures
\subsection{Discussion and Result}
\subsubsection{FRDSSC $\&$ IMI}
With respect to equation (6) and $(N=60, \hspace{0.1cm}\sigma=3.3)$, we will have
\begin{equation}\label{46}
F_1=0.00141679,\hspace{0.5cm}  F_{2}=-0.0175289
\end{equation}
Given the inflation model and the refined de Sitter swampland conjecture (RDSSC), we have the following information: The constants \(c_1\) and \(c_2\) are related to the  \(F_1\) and \(F_2\). The RDSSC requires that \(c_1\) and \(c_2\) should not be of order \(\mathcal{O}(1)\). Now, We evaluate this model using the FRDSSC. From equations (5), we will have,
\begin{equation}\label{47}
(0.00141679)^q+a (0.0175289)\geq1-a,
\end{equation}
By selecting q=2.2, we will have
\begin{equation}\label{48}
a\geq 0.982773
\end{equation}
Therefore, if we consider $a=0.982873$,  b is equal to 0.017127. So IMI is compatible with the FRDSSC. Similarly, if we also assume $(N=55,\hspace{0.1cm}\sigma=3.3)$ and $(N=70,\hspace{0.1cm}\sigma=3.3)$, our model will violate (RDSSC). But by using the manual adjustment that we consider for (FRDSSC) with respect to its free parameters (a, b,q), the conjecture can be satisfied by our inflationary model. Therefore, for different values assumed for $(N,\hspace{0.1cm}\sigma)$, our model will generally violate (RDSSC) and satisfy (FRDSSC)
\subsubsection{SWGC, SSWGC $\&$ IMI}
We obtain the potential derivatives (22) to study other conjectures such as SWGC and SSWGC. According to the potential (22) and its derivatives, we have the following expression for SWGC,
\begin{equation}\label{48}
\sigma^2 (1 + \sigma)^2 M^{8 + 2 \sigma} \phi^{-2 (3 + \sigma)} [(\sigma+2)^2- \phi^2]\geq 0
\end{equation}
When $\phi\leq \sigma+2$, the above relationship is met. In this case, SWGC will be satisfied with the IMI.
Also, we can evaluate the SSWGC with respect to potential (22) and its derivatives. So, we have,
\begin{equation}\label{49}
\sigma^2 (1 + \sigma)^2 M^{8 + 2 \sigma} \phi^{-2 (3 + \sigma)} [(\sigma+2)(\sigma+1)- \phi^2]\geq 0
\end{equation}
When $\phi \leq \sqrt{(\sigma+2)(\sigma+1)}$, we can find the points that the SSWGC holds for our model.\\
\begin{center}
\begin{table}
  \centering
\begin{tabular}{|p{1.5cm}|p{1.5cm}||p{2.5cm}|p{2.5cm}||p{2.9cm}|}
  \hline
  The Model  & (RDSSC) & FRDSSC & SWGC & SSWGC\\[3mm]
   \hline
  \vspace{0.2cm}$\frac{M^{4+\sigma}}{\phi^{\sigma}}$ & \hspace{1cm} $Violated$ & \hspace{2cm} Satisfied (a=0.982873,\hspace{0.1cm} b=0.017127,\hspace{0.1cm} q=2.2) & \hspace{2cm}Satisfied ($\phi\leq \sigma+2$)& \hspace{2cm} \hspace{-0.4cm}Satisfied ($\phi \leq \sqrt{(\sigma+2)(\sigma+1)}$)  \\[3mm]
   \hline
\end{tabular}
\caption{Summary of the results.}\label{1}
\end{table}
 \end{center}
\newpage
\section{conclusions}
So far, inflationary models have been studied from different perspectives and using various conditions such as slow-roll, constant-roll, and the conjectures of the swampland program. In this paper, we investigated the inverse monomial inflation (IMI) on the brane. We limited our results to potentials of the form \(\frac{M^{4+\sigma}}{\phi^{\sigma}}\), where \(M\) and \(\sigma\) are constants. We calculated several cosmological parameters, including the spectral index \(n_S\), the tensor-to-scalar ratio \(r\), and the running of the spectral index \(\alpha_s\).
Next, we examined the model's compatibility with several swampland conjectures. Specifically, we checked its alignment with the refined de Sitter swampland conjecture (RDSSC), the further refining de Sitter swampland conjecture (FRDSSC), the scalar weak gravity conjecture (SWGC), and the strong scalar weak gravity conjecture (SSWGC). Despite the model's incompatibility with the RDSSC, we identified a specific region where it is compatible with the other conjectures. For instance, the model satisfies the FRDSSC with parameters \(a = 0.982873\), \(b = 0.017127\), and \(q = 2.2\). Additionally, the model is compatible with the SWGC under the condition \(\phi \leq \sigma + 2\) and consistent with the SSWGC under the condition \(\phi \leq \sqrt{(\sigma + 2)(\sigma + 1)}\). We summarized these results in Table 1.
Inverse monomial inflation (IMI) on the brane emerges as a promising candidate for a realistic inflation model of the universe, as it can satisfy several swampland conjectures simultaneously. The swampland conjectures are a set of criteria that distinguish effective field theories that can be consistently coupled to quantum gravity from those that cannot. Therefore, it is crucial to investigate how to derive IMI from string theory, which is a leading framework for quantum gravity.
Moreover, we should explore other inflation models that can also meet these swampland criteria and compare them with IMI. By doing so, we may discover common features of inflation models that are compatible with quantum gravity. This comparative analysis could provide deeper insights into the nature of inflation and its connection to fundamental theories of physics. Additionally, understanding these common features could guide the development of new models that are both theoretically sound and observationally viable.
In conclusion, our study highlights the potential of IMI on the brane as a viable inflationary model that aligns with several swampland conjectures. This alignment not only supports the model's theoretical foundation but also enhances its credibility as a candidate for describing the early universe.
However, the non-canonical scalar field and potentials are also a very powerful inflation framework\cite{666,6666,66666}. Future research should focus on deriving IMI from string theory and exploring other models that satisfy the swampland criteria, thereby contributing to the broader understanding of inflation and its role in the context of quantum gravity.\\


\begin{thebibliography}{11}
\bibitem{101}
D. Harlow, B. Heidenreich, M. Reece, and T. Rudelius, "The Weak Gravity Conjecture: A Review", arXiv:2201.08380 (2022).
\bibitem{102}
J. Sadeghi, B. Pourhassan, S. N. Gashti, and S. Upadhyay, "Weak Gravity Conjecture, Black Branes and Violations of Universal Thermodynamic Relation", Annals of Physics 447, (1), 169168 (2022).
\bibitem{103}
C. Cheung and Grant N. Remmen, "Naturalness and the Weak Gravity Conjecture", Phys. Rev. Lett. 113, 051601 (2014).
\bibitem{104}
J. Sadeghi, S. Noori Gashti, I. Sakalli, and B. Pourhassan, "Weak Gravity Conjecture of Charged-Rotating-AdS Black Hole Surrounded by Quintessence and String Cloud", NPB (2023).
\bibitem{105}
Y. Hamada, T. Noumi, and G. Shiu, "Weak Gravity Conjecture from Unitarity and Causality",
Phys. Rev. Lett. 123, 051601 (2019).
\bibitem{106}
J. Sadeghi, M. Shokri, M. R. Alipour, and S. Noori Gashti, "Weak Gravity Conjecture from Conformal Field Theory: A Challenge from Hyperscaling Violating and Kerr-Newman-AdS Black Holes", Chinese Physics C 47 (1), 015103 (2022).
\bibitem{107}
L. Ma, Y. Pang, and H. Lü, "$\alpha$'-corrections to near extremal dyonic strings and weak gravity conjecture", Journal of High Energy Physics 2022, 157 (2022).
\bibitem{108}
J. Sadeghi, M. R. Alipour, S. N. Gashti, "Strong Cosmic Censorship in light of Weak Gravity Conjecture for Charged Black Holes", Journal of High Energy Physics 2023 (2), 1-14 (2023).
\bibitem{108'}
J. Sadeghi, B. Pourhassan, S. N. Gashti, I. Sakallı, M. R. Alipour, "de Sitter Swampland Conjecture in String Field Inflation", The European Physical Journal C 83 (635), 2023 (2023).
\bibitem{777}
Hossain, Md Wali, et al. "Variable gravity: A suitable framework for quintessential inflation." Physical Review D 90.2 (2014): 023512.
\bibitem{7777}
Sheikhahmadi, Haidar, et al. "Hamilton-Jacobi formalism for inflation with non-minimal derivative coupling." Journal of Cosmology and Astroparticle Physics 2016.10 (2016): 021.
\bibitem{77777}
Karydas, Stelios, Eleftherios Papantonopoulos, and Emmanuel N. Saridakis. "Successful Higgs inflation from combined nonminimal and derivative couplings." Physical Review D 104.2 (2021): 023530.
\bibitem{1}
E. Palti, "The swampland: introduction and review", Fortsch. Phys. 67, 6, 1900037 (2019).
\bibitem{1'}
J. Sadeghi, B. Pourhassan, S. Noori Gashti, S. Upadhyay, E. Naghd Mezerji, "The emergence of universal relations in the AdS black holes thermodynamics", Physica Scripta 98 (2), 025305 (2023).
\bibitem{2}
N. A. Hamed, L. Motl, and A. Nicolis, "The string landscape, black holes and gravity as the weakest force", JHEP 0706, 060 (2007).
\bibitem{2'}
S. N. Gashti, J. Sadeghi, B. Pourhassan, "Pleasant behavior of swampland conjectures in the face of specific inflationary models", Astroparticle Physics 139, 102703 (2022).
\bibitem{3}
Y. Akrami, R. Kallosh, A. Linde, and V. Vardanyan, "The landscape, the swampland and the era of precision cosmology", Fortsch. Phys. 67, 1-2, 1800075 (2019).
\bibitem{4}
T. Brennan, F.Carta, and C. Vafa, "The string landscape, the swampland, and the missing corner", PoS TASI2017, 015 (2017).
\bibitem{5}
H. Murayama, M. Yamazaki, and T. Yanagida, "Do we live in the swampland?", JHEP 12, 032 (2018).
\bibitem{6}
C. Vafa, "The string landscape and the swampland". arXiv hep-the/0509212 (2005).
\bibitem{7}
E. Palti, "The weak gravity conjecture and scalar fields", J. High Energy Phys, 8, 034 (2017).
\bibitem{8}
K. Kooner, S. Parameswaran, and I. Zavala, "Warping the weak gravity conjecture", Phys. Lett. B 759, 402409 (2016).
\bibitem{9}
M. Montero, G. Shiu, and P. Soler, "The weak gravity conjecture in three dimensions", JHEB 2016, 159 (2016).
\bibitem{10}
P. Saraswat, "Weak gravity conjecture and effective field theory", Phys. Rev. D 95, 025013 (2017).
\bibitem{11}
Y. Akayama, and Y. Nomura, "Weak gravity conjecture in the AdS/CFT correspondence", Phys. Rev. D 92, 126006 (2015).
\bibitem{12}
J. Sadeghi, S. Noori Gashti, and E. Naghd Mezerji, "The investigation of universal relation between corrections to entropy and extremality bounds with verification WGC", Phys. Dark Univ 30, 100626 (2020).
\bibitem{109}
J. Sadeghi, E. N. Mezerji, and S. N. Gashti, "Study of some cosmological parameters in logarithmic corrected  gravitational model with swampland conjectures", Modern Physics Letters A 36, (05), 2150027 (2021).
\bibitem{110}
J. Sadeghi, and S. N. Gashti, "Anisotropic constant-roll inflation with noncommutative model and swampland conjectures", The European Physical Journal C 81, 1-10, (2021).
\bibitem{111}
M. van Beest, J. Calderón-Infante, D. Mirfendereski, and I. Valenzuela, "Lectures on the swampland program in string compactifications". Physics Reports, 989, 1 (2022).
\bibitem{112}
S. N. Gashti, J. Sadeghi, and M. R. Alipour, "Further Refining Swampland dS Conjecture in Mimetic f (G) Gravity", IJMPD (2023).
\bibitem{113}
D. Andriot, and C.  Roupec, "Further Refining the de Sitter Swampland Conjecture", Fortschritte Phys. 67, 1800105 (2019).
\bibitem{114}
S. N. Gashti, "Two-field inflationary model and swampland de Sitter conjecture",
Journal of Holography Applications in Physics 2, (1), 13-24 (2022).
\bibitem{115}
J. Sadeghi, B. Pourhassan, S. N. Gashti, and S. Upadhyay, "Swampland conjecture and inflation model from brane perspective,  Physica Scripta 96, (12), 125317 (2021).
\bibitem{116}
H. Ooguri, E. Palti, G. Shiu, and C. Vafa, "Distance and de Sitter conjectures on the Swampland", Phys. Lett. 788, 180 (2019).
\bibitem{117}
J. Sadeghi, B. Pourhassan, S. Noori Gashti, E. Naghd Mezerji, and A. Pasqua, "Cosmic Evolution of the Logarithmic f(R) Model and the dS Swampland Conjecture", Universe 8, (12), 623 (2022).
\bibitem{118}
J. Sadeghi, S. Noori Gashti, and M. R. Alipour, "Notes on further refining de Sitter swampland conjecture with inflationary models", Chinese Journal of Physics 79, 490 (2022).
\bibitem{119}
S. N. Gashti, J. Sadeghi, S. Upadhyay, and M. R. Alipour, "Swampland dS conjecture in mimetic f (R, T) gravity", Communications in Theoretical Physics 74, (8), 085402 (2022).
\bibitem{120}
Y. Liu, "Higgs inflation and its extensions and the further refining dS swampland conjecture", Eur. Phys. J. C 82, 1052 (2022).
\bibitem{121}
S. N. Gashti, and J. Sadeghi, "Refined swampland conjecture in warm vector hybrid inflationary scenario",
The European Physical Journal Plus 137, (6), 1-13 (2022).
\bibitem{122}
Y. Liu, "Higgs inflation and scalar weak gravity conjecture", Eur. Phys. J. C 81, 1122 (2022).
\bibitem{123}
J. Sadeghi, M. R. Alipour, and S. Noori Gashti, "Scalar Weak Gravity Conjecture in Super Yang-Mills Inflationary Model", Universe 8, 621 (2022).
\bibitem{124}
E. Gonzalo, and L. Ibáñez, "A Strong Scalar Weak Gravity Conjecture and some implications, J. High Energy Phys. 8, 118 (2019).
\bibitem{125}
J. Sadeghi, and S. N. Gashti, "Investigating the logarithmic form of f(R) gravity model from brane perspective and swampland criteria", Pramana 95 (198) (2022).
\bibitem{126}
J. Yuennan, P. Channuie, "Composite Inflation and further refining dS swampland conjecture", Nuclear Physics B 986, 116033 (2023).
\bibitem{127}
J. Sadeghi, S. N. Gashti, and F. Darabi, "Swampland conjectures in hybrid metric-Palatini gravity,
Physics of the Dark Universe", 101090 (2022).
\bibitem{13}
H. Ooguri, C.  Vafa, "On the Geometry of the String Landscape and the Swampland", Nucl. Phys. B, 766, 21 (2007).
\bibitem{128}
S. N. Gashti, J. Sadeghi, "Constraints on cosmological parameters in light of the scalar–tensor theory of gravity and swampland conjectures",  Modern Physics Letters A 37, (18), 2250110 (2022).
\bibitem{129}
S. Noori Gashti, and J. Sadeghi, "Inflation, swampland and landscape", International Journal of Modern Physics A 37 (04), 2250006 (2022).
\bibitem{14}
G. Obied, H. Ooguri, L. Spodyneiko, and C. Vafa, "de Sitter Space and the Swampland", arXiv:1806.08362  (2018).
\bibitem{15}
W. H. Kinney, S. Vagnozzi, and L. Visinelli, "The zoo plot meets the swampland: mutual (in) consistency of single-field inflation, string conjectures, and cosmological data",Class. Quant. Grav. 36, 117001 (2019).
\bibitem{16}
A. Achcarro, and G. A. Palma, "The string swampland constraints require multi-field inflation",  JCAP 02, 041 (2019).
\bibitem{17}
S. K. Garg, and C. Krishnan, "Bounds on slow roll and the de Sitter swampland", JHEP 11  075 (2019).
\bibitem{18}
H. Ooguri, and C. Vafa, "Non-supersymmetric AdS and the Swampland", Adv. Theor. Math. Phys. 21, 1787 (2017).
\bibitem{19}
P. Agrawal, G. Obied, P. J. Steinhardt, and C. Vafa, "On the cosmological implications of the string swampland", Phys. Lett. B 784, 271 (2018).
\bibitem{20}
A. Kehagias, and A. Riotto, "A note on Inflation and the Swampland", Fortsch.Phys. 66 10, 1800052 (2018) .
\bibitem{21}
S. Brahma, and M. Wali Hossain, "Avoiding the string swampland in single-field inflation: Excited initial states", JHEP 03, 006 (2019) .
\bibitem{22}
S. Das, "Note on single-field inflation and the swampland criteria", Phys. Rev. D 99, 8, 083510 (2019).
\bibitem{5000}
Sasaki, Misao, Tetsuya Shiromizu, and Kei-ichi Maeda. "Gravity, stability, and energy conservation on the Randall-Sundrum brane world." Physical Review D 62.2 (2000): 024008.
\bibitem{5001}
Bowcock, Peter, Christos Charmousis, and Ruth Gregory. "General brane cosmologies and their global spacetime structure." Classical and Quantum Gravity 17.22 (2000): 4745.
\bibitem{5000'}
Apostolopoulos, Pantelis S., et al. "Mirage effects on the brane." Physical Review D—Particles, Fields, Gravitation, and Cosmology 72.4 (2005): 044013.
\bibitem{5000''}
Diakonos, F. K., and E. N. Saridakis. "Statistical approach to the cosmological-constant problem on brane worlds." Journal of Cosmology and Astroparticle Physics 2009.02 (2009): 030.
\bibitem{5000'''}
Odintsov, S. D., V. K. Oikonomou, and Emmanuel N. Saridakis. "Superbounce and loop quantum ekpyrotic cosmologies from modified gravity: F (R), F (G) and F (T) theories." Annals of Physics 363 (2015): 141-163.
\bibitem{23}
J. M. Cline, C. Grojean, and G. Servant, "Cosmological expansion in the presence of an extra dimension", Phys. Rev. Lett. 83, 4245 (1999).
\bibitem{24}
C. Csaki, M. Graesser, C. F. Kolda, and J. Terning, "Cosmology of one extra dimension with localized gravity", Phys. Lett. B 462, 34 (1999).
\bibitem{25}
P. Binetruy, C. Deffayet, U. Ellwanger, and D. Langlois, "Brane cosmological evolution in a bulk with cosmological constant", Phys. Lett. B 477, 285 (2000).
\bibitem{26}
K. Freese, and M. Lewis, "Cardassian expansion: a model in which the universe is flat, matter dominated, and accelerating", Phys. Lett. B 540, 1 (2002).
\bibitem{27}
R. Maartens, D. Wands, B. A. Bassett, and I. Heard, "Chaotic inflation on the brane", Phys. Rev. D 62, 041301 (2000).
\bibitem{28}
M. Jerome, R. Christophe, and V. Vincent, "Encyclopædia Inflationaris", Phys. Dark Univ. 5-6, 75--235 (2014).
\bibitem{29}
D. Langlois, R. Maartens, and D. Wands, "Gravitational waves from inflation on the brane", Phys. Lett. B 489, 259 (2000).
\bibitem{30}
M. C. Bento, R. G. Felipe, and N. M. C. Santos, "Brane assisted quintessential inflation with transient acceleration", Phys. Rev. D 77, 123512 (2008).
\bibitem{300}
H. Es-sobbahi, and M. Nach, "On braneworld inverse power-law inflation", International Journal of Modern Physics A. 33, 10, 1850058 (2018).
\bibitem{1000}
Ichiki, K., et al. "Observational constraints on dark radiation in brane cosmology." Physical Review D 66.4 (2002): 043521.
\bibitem{1001}
Sasankan, Nishanth, et al. "New observational limits on dark radiation in braneworld cosmology." Physical Review D 95.8 (2017): 083516.
\bibitem{1002}
Maartens, Roy, et al. "Chaotic inflation on the brane." Physical Review D 62.4 (2000): 041301.
\bibitem{31}
Y. Akrami et al, "Planck 2018 results-X. Constraints on inflation", A$\&$A 641, A10 (2020).
\bibitem{a}
G. Huey and J. E. Lidsey, "Inflation, braneworlds and quintessence", Phys. Lett. B 514, 217 (2001).
\bibitem{b}
E. J. Copeland, A. R. Liddle and J. E. Lidsey, "Steep inflation: ending braneworld inflation by gravitational particle production", Phys. Rev. D 64, 023509 (2001).
\bibitem{b'}
Huey, Greg, and James E. Lidsey. "Inflation, braneworlds and quintessence." Physics Letters B 514.3-4 (2001): 217-225.
\bibitem{b''}
Copeland, Edmund J., Andrew R. Liddle, and James E. Lidsey. "Steep inflation: Ending braneworld inflation by gravitational particle production." Physical Review D 64.2 (2001): 023509.
\bibitem{c}
L. R. Abramo and F. Finelli, "Cosmological dynamics of the tachyon with an inverse power-law potential", Phys. Lett. B 575, 165 (2003).
\bibitem{666}
Geng, Chao-Qiang, et al. "Quintessential inflation with canonical and noncanonical scalar fields and Planck 2015 results." Physical Review D 92.2 (2015): 023522.
\bibitem{6666}
Geng, Chao-Qiang, et al. "Observational constraints on successful model of quintessential Inflation." Journal of Cosmology and Astroparticle Physics 2017.06 (2017): 011.
\bibitem{66666}
Lola, Smaragda, Andreas Lymperis, and Emmanuel N. Saridakis. "Inflation with non-canonical scalar fields revisited." The European Physical Journal C 81.8 (2021): 719.
\end{thebibliography}
\end{document}